\def\year{2024}\relax
\documentclass[letterpaper]{article} 
\usepackage{aaai22}  
\usepackage{times}  
\usepackage{helvet}  
\usepackage{courier}  
\usepackage[hyphens]{url}  
\usepackage{graphicx} 
\usepackage{natbib}  
\usepackage{caption} 
\DeclareCaptionStyle{ruled}{labelfont=normalfont,labelsep=colon,strut=off} 
\usepackage{algorithm}
\usepackage{algorithmic}

\usepackage{xcolor}
\usepackage{newfloat}
\usepackage{listings}
\floatstyle{ruled}
\newfloat{listing}{tb}{lst}{}
\floatname{listing}{Listing}

\title{The Peripatetic Hater: Predicting Movement Among Hate Subreddits}
 \author{Daniel Hickey\textsuperscript{\rm 1},
 Daniel M.T. Fessler\textsuperscript{\rm 2,3,4},
 Kristina Lerman\textsuperscript{\rm 5},
 Keith Burghardt\textsuperscript{\rm 5}
 }
 \affiliations {
 \textsuperscript{\rm 1}School of Information, University of California, Berkeley, Berkeley, CA 94704, USA\\
 \textsuperscript{\rm 2}Department of Anthropology, University of California, Los Angeles, Los Angeles, CA 90095, USA\\
 \textsuperscript{\rm 3}Bedari Kindness Institute, University of California, Los Angeles, Los Angeles, CA 90095, USA\\
 \textsuperscript{\rm 4}Center for Behavior, Evolution, \& Culture, University of California, Los Angeles, Los Angeles, CA 90095, USA\\
 \textsuperscript{\rm 5}USC Information Sciences Institute, Marina del Rey, CA 90292, USA\\
 dan\_hickey@berkeley.edu, dfessler@anthro.ucla.edu, lerman@isi.edu, keithab@isi.edu
 }

\begin{document}

\maketitle

\begin{abstract}
Many online hate groups exist to disparage others based on race, gender identity, sex, or other characteristics. The accessibility of these communities allows users to join multiple types of hate groups (e.g., a racist community and a misogynistic community), raising the question of whether users who join additional types of hate communities could be further radicalized compared to users who stay in one type of hate group. However, little is known about the dynamics of joining multiple types of hate groups, nor the effect of these groups on peripatetic users. We develop a new method to classify hate subreddits and the identities they disparage, then apply it to understand better how users come to join different types of hate subreddits. The hate classification technique utilizes human-validated deep learning models to extract the protected identities attacked, if any, across 168 subreddits. We find distinct clusters of subreddits targeting various identities, such as racist subreddits, xenophobic subreddits, and transphobic subreddits. We show that when users become active in their first hate subreddit, they have a high likelihood of becoming active in additional hate subreddits of a different category. We also find that users who join additional hate subreddits, especially those of a different category develop a wider hate group lexicon. These results then lead us to train a deep learning model that, as we demonstrate, usefully predicts the hate categories in which users will become active based on post text replied to and written. The accuracy of this model may be partly driven by peripatetic users often using the language of hate subreddits they eventually join. Overall, these results highlight the unique risks associated with hate communities on a social media platform, as discussion of alternative targets of hate may lead users to target more protected identities.
\end{abstract}

\section{Introduction}
Recently, there has been extensive work documenting online hate communities and the impact that they have on the users of social media platforms \cite{schmitz2023users, russo2023spillover}. Such research is motivated by the combination of substantial increases in hate crimes \cite{FBIhate}, and the observation that radicalized individuals attribute their beliefs to online sources \cite{start}. Much of the existing literature in this domain provides either case studies of individual communities \cite{chandrasekharan2017, schmitz2023users,russo2023spillover} or broad analyses of groups centered around a common theme, such as the manosphere \cite{ribeiro2021evolution}. Additionally, although prior work observes strong user overlap between anti-feminism and alt-right communities \cite{mamie2021anti}, such research does not show how the core themes discussed in these communities are influenced by the users active in both communities. 

Here, we extend the existing research corpus by curating a comprehensive dataset of 168 hate subreddits and using deep learning models and clustering techniques to group them by ideology (racist, misogynistic, xenophobic, transphobic, homophobic, Islamophobic, antisemitic, ableist, and general hate). In these subreddits, we analyze \textit{peripatetic users} who participate in multiple categories of hate subreddits, attempting to characterize them and understand the impact that they have on Reddit's ecosystem of hate. Our research questions are as follows:

\begin{enumerate}
    \item[\textbf{RQ1}] Are users who participate in one hate community likely to participate in hate communities of other types?
    \item[\textbf{RQ2}]  What characteristics distinguish peripatetic and non-peripatetic users?
    \item[\textbf{RQ3}]How might joining additional hate communities affect user behavior?
    \item[\textbf{RQ4}]  Given that a user has already participated in a hate community, can we predict the types of hate communities in which they will participate in the future?
\end{enumerate}

Our findings help answer each respective research question: (RQ1) Using matched-pair analysis, we find that participation in a given hate subreddit is associated with joining other hate subreddits at a rate roughly \emph{two times higher} than that of matched users. (RQ2) Using BERTopic \cite{grootendorst2022bertopic} and lexicons of ingroup language, we find that peripatetic users introduce into their initial subreddit topics of discussion from the categories in which they eventually participate. For example, a user who initially posts only in a misogynistic subreddit, but subsequently also posts in a racist subreddit, is more likely to introduce language characteristic of racist subreddits into the misogynistic subreddit. (RQ3) We find that users who join additional hate subreddits become more active in hate subreddits as a whole and use lexicons associated with these newly joined subreddits. Both results are more pronounced when users join hate subreddits of a category that differs from that of the first subreddit they joined. (RQ4) Finally, we develop a deep-learning model to predict the subreddit categories these peripatetic users will join. This task is challenging, as we are differentiating among users who already exhibit signs of hostility. We find that the language of users who eventually become peripatetic is strongly associated with the subreddit categories that they subsequently join. 

Overall, our results offer a novel perspective on how hate communities may converge and become more ``general'' in an online environment as a result of peripatetic user activity. All data and code are in the following repository \url{https://anonymous.4open.science/r/peripatetic-hater-1D26}.

\begin{figure}[ht]
    \centering
    \includegraphics[width=\columnwidth]{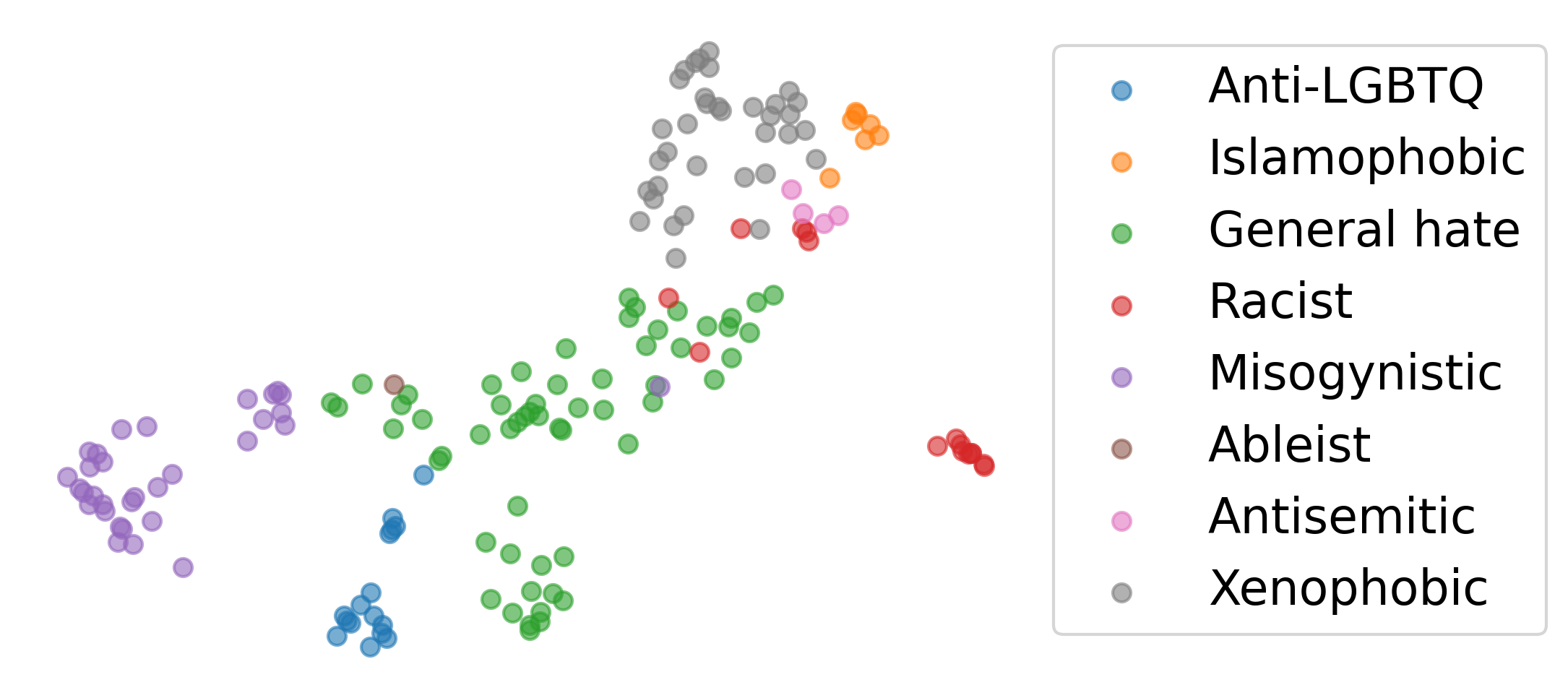}
    \caption{UMAP plot of hate speech distributions for each subreddit. Each point represents a subreddit, and their color corresponds with the K-means cluster they have been assigned to.}
    \label{fig:UMAP}
\end{figure}

\section{Related Work}
\subsection{Types of Online Hate Communities}
We begin our exploration of online hate communities by building on other scholars' prior research into specific categories of such communities; below we outline these categories and review work that has examined them.
\textbf{Misogynist.} Extensive work has investigated the various identities targeted by online hate communities. One of the most widely studied groups is the manosphere, a collection of communities centered around misogyny and anti-feminism \cite{marwick2018drinking, habib2022making}. Despite sharing this common theme, there is a wide diversity of ideologies within the manosphere. Incels, for example, believe they are incapable of finding a sexual or romantic partner, and blame women and society for their misfortune \cite{daly2022think}. This is in contrast with pick-up artists, who discuss strategies for seducing women, and ``men going their own way'' (MGTOW), who voluntarily decide to be celibate as they believe they should distance themselves from women \cite{lin2017antifeminism}. A recent analysis shows MGTOW and incel communities are growing in popularity and are more toxic than other manosphere communities, which are decreasing in popularity \cite{ribeiro2021evolution}. Critically, members of incel communities have engaged in mass shooting events \cite{barcellona2022incel}.

\textbf{Anti-LGBTQ.} Anti-LGBTQ communities are also prevalent on the internet, including trans-exclusionary radical feminists (TERFs), an anti-transgender group. TERFs generally subscribe to a binary view of gender that is aligned with biological sex (disregarding intersex individuals), and see transgender people as a threat to women's spaces \cite{jones2020toilet}. Scholars have previously demonstrated flaws in such discourse and the harm it produces \cite{williams2020ontological}. TERF communities have been present on Reddit in the past, with r/GenderCritical being the most prominent \cite{tiffany2020terfs}. Previous work has found these communities to be more toxic than mainstream feminist communities \cite{balci2023beyond}, and models have been built to automatically identify the discourse employed by them \cite{lu2022subtle}.

\textbf{Racist/xenophobic.} Racist groups have been highly prevalent on the internet, and on Reddit in particular \cite{chandrasekharan2017}, as have a class of online groups known collectively as the alt-right, an extreme political movement that gained popularity through its utilization of the unique characteristics of social media \cite{winter2019online}. The alt-right believes that the cultural identity of white individuals is at risk, with movements for social justice and political correctness being seen as principal threats \cite{splc_altright}. Alt-right online communities have radicalized users in part through the use of memes \cite{dafaure2020great}, with one analysis showing that communities linked to the alt-right have a strong influence on the spread of memes in the larger web ecosystem \cite{zannettou2018origins}. The misogynist movement ``GamerGate'' helped the alt-right movement gain adherents \cite{bezio2018ctrl}, demonstrating a relationship between seemingly distinct types of hate.

\subsection{Radicalization Pathways and Gateway Communities}

Radicalization is explicable in terms of a number of sociological and psychological effects \cite{borum2011radicalization,maskaliunaite2015exploring,neo2016internet}, which include: uncertainty reduction \cite{rootsofextremism}, achieving significance \cite{kruglanski2014psychology}, and the need to be popular \cite{siegel2011dying}. A recent study theorized that many users initially participate in extreme political subreddits as they are disillusioned with a polarized political landscape and desire to hold beliefs outside of mainstream political viewpoints \cite{mann2023unsorted}. Additionally, a study that found highly-upvoted posts on r/The\_Donald were likely to contain anti-Muslim sentiment argued that the high number of upvotes normalized such viewpoints, strengthening the bonds of the ingroup as they united against an ``other.'' \cite{gaudette2021upvoting}.  
A substantial corpus of existing research examines the micro-scale processes that drive people to radicalization. At the level of categories of online groups, scholars have studied whether certain communities are ``gateways'' to more extreme communities, such as the connection between the intellectual dark web and the alt-right \cite{ribeiro2020auditing}, or between the manosphere and the alt-right \cite{ribeiro2020auditing, barcellona2022incel}. Other studies measure traits indicative of user radicalization or extremism that are adopted by members of certain hate groups, finding that such traits increase once users participate in hate subreddits \cite{habib2022making, schmitz2023users}. Recommendation algorithms meanwhile may not necessarily contribute significantly to radicalization as users who consume extreme content on YouTube are driven to do so by external sources, and are characterized by high racial and gender resentment even before engaging with such content \cite{chen2023subscriptions}.

\subsection{Movement Among Online Communities}


A growing body of literature is devoted to understanding how users participate in multiple communities on the same platform. One theme in this research is the attempt to understand inter-community conflict, where members active in one community make posts in other communities in order to attack them \cite{efstratiou2023non, datta2019extracting}. Some authors speculate that mass migrations of users among communities on Reddit may disrupt the linguistic evolution of destination communities, potentially influencing their likelihood of being banned \cite{habib2021proactive}. More generally, Rollo et al. \cite{rollo2022communities} devise a model of ``attention flow'' to quantitatively describe how users on Reddit shift their focus among subreddits over time. Scholars have also built models to predict whether users will participate in particular online communities \cite{jin2010predicting, debaere2018multi}, with, for example, one study finding that the social support that users in an online health community receive is predictive of their remaining in that community \cite{wang2017analyzing}.

The present study builds on previous research by examining a multitude of hate communities while assessing the connections among all of them. In addition, we describe the tendency of users to subsequently join specific additional hate communities after initially joining a given hate community, and measure the user behavior that predates such additional participation. Finally, we build a model to predict which new hate subreddits users will join.

\section{Methods}

\subsection{Collecting a Dataset of Hate Subreddits}

To properly understand how different categories of hate subreddits influence each other, we define a set of hate subreddits that promulgate hate for a variety of identities. While researchers have previously curated exhaustive lists of subreddits that represent a particular ideology, such as the manosphere \cite{ribeiro2021evolution} or the alt-right \cite{mamie2021anti}, there currently exists no comprehensive dataset of hate subreddits that captures the diversity of ideologies that can be observed on the platform. We therefore create a large dataset of hate subreddits, each labeled by the identity or identities that it targets. Our process consists of first collecting a wide set of subreddits that are likely to be hate subreddits, detecting the targets of hate in each subreddit, and clustering subreddits together that contain hate speech of similar types.

To identify potential hate subreddits, we collect curated lists from various sources. First, we find lists of hate subreddits or banned subreddits previously identified in the literature \cite{mamie2021anti, ribeiro2021evolution, vidgen2021introducing, schmitz2023users}. As these papers cite subreddits such as r/AgainstHateSubreddits to find potential hate groups, we supplement these lists with more subreddits by searching ``list of hate subreddits'' on r/AgainstHateSubreddits and finding each post listing hate subreddits within the first 100 search results as of October 2023. We then collect the entire history of comments and submissions from each potential hate subreddit using the Pushshift API \cite{Baumgartner2020}. All subreddits with fewer than 1,000 users are removed from further analysis in order to have sufficient data for each subreddit, this choice removes less than 2\% of all users in the dataset. Additionally, as many banned subreddits could be unrelated to hate (such as subreddits concerned with the black market), one annotator (who is also an author of this paper) viewed 10 random comments and 10 random submissions from each subreddit, filtering out subreddits that were clearly unrelated to hate. At the end of this process, 168 subreddits remained. The links to all subreddits analyzed, as well as the finalized list of subreddits, are available in the repository. 

We next define identities targeted in each hate subreddit based on groups identified in United States federal hate crime law \footnote{\url{https://www.justice.gov/hatecrimes/learn-about-hate-crimes}, accessed 15 Sep, 2024}, where a hate crime is defined as ``a crime motivated by bias against race, color, religion, national origin, sexual orientation, gender, gender identity, or disability.'' We then further divide the ``religion'' category into two categories, representing hate speech directed at Jews and hate speech directed at Muslims, respectively, as hate speech directed at these groups often additionally targets their race or national origin. Our final list of hate speech categories thus consists of antisemitism, Islamophobia, ableism, misogyny, xenophobia, racism, homophobia, and transphobia. 

To find the prevalence of each category targeted in each subreddit, we train a deep learning model using the Measuring Hate Speech corpus \cite{sachdeva-etal-2022-measuring}. The Measuring Hate Speech corpus consists of 50k posts collected from X, YouTube, and Reddit. Each post has been coded by multiple annotators as to the identities targeted by the post, as well as whether the post constitutes hate speech (under several different levels/definitions of hate speech). To train our prediction model, we use the Demux architecture, a multi-label transformer-based architecture that generates embeddings for the names of labels, leveraging associations between related labels to improve prediction performance \cite{chochlakis2023leveraging}. The names of the labels are prepended to each input sequence, and the text of the posts and labels are embedded using BERT-base. We note that after analyzing 50 posts per subreddit using the Google Translate API \footnote{\url{https://cloud.google.com/translate}, accessed 15 Sep, 2024}, the average percentage of English posts per subreddit is 97\%, implying an English-language BERT is appropriate.

These embeddings are input into a two-layer neural network with the sigmoid activation function applied to generate predictions. The model is fit using the binary cross-entropy loss function. We use a hidden layer size of 128, a dropout value of 0.1, and the Tanh activation function in the hidden layer based on similar hyperparameters that achieve high performance on this exact dataset \cite{sachdeva2022targeted}, and we noticed minimal changes in performance when testing other values for the hyperparameters. As a baseline, we also train a basic BERT model with the same hyperparameters and compare the performance of the two models. We train both models using an NVIDIA RTX A6000 GPU.

Given that the ``targeted identity'' labels in the Measuring Hate Speech corpus are only intended to capture whether an identity group is \emph{mentioned} by a post and not whether that post constitutes hate speech against the identity group \cite{sachdeva2022targeted}, we combine a model to predict identity group label with other labels from the measuring hate speech corpus (negative sentiment, disrespect, insults, attacks on identity groups, and a canonical ``hate speech'' label). Based on the annotations described below, we found that the best performance was obtained using a model that defines hate against a targeted identity as one that detects both negative sentiment and a targeted identity in the text. All reports of model performance and analyses are therefore generated using this classifier. Other labels (i.e., disrespect, etc) yield similar results. To illustrate that these classifiers are similar, we display the matrix of Spearman correlations among the predicted probabilities of these classifiers in Table~\ref{tab:correlation_matrix}.

To validate our method of detecting hate speech, three human annotators, fluent in English, labeled 10 comments and 10 submissions each from a set of 25 random hate subreddits. The annotators assessed the number of comments and submissions containing hate speech in each category for each subreddit. To determine whether a given post constitutes hate speech, the annotators referred to the United Nations' definition of hate speech\footnote{\url{https://www.un.org/en/hate-speech/understanding-hate-speech/what-is-hate-speech}, accessed 15 Sep, 2024}. The annotators achieved a Krippendorff's alpha score of 0.52 \cite{hayes2007answering}, implying relatively poor inter-rater reliability, consistent with other papers on hate speech \cite{vidgen-etal-2021-introducing,sachdeva-etal-2022-measuring}, and indicative of the difficulty to assess identities targeted. For each category, we calculate the $R^2$ value between the average values of human annotations and the number of posts predicted as hate speech for that category by Demux. Demux confidence thresholds for negative sentiment and identity targeted are varied to maximize the F1 score of predictions on the test set of the Measuring Hate Speech corpus. Table~\ref{tab:mhs_performance}. To obtain estimates of uncertainty for the performance of our classifier, we train the model five separate times, using a different random split of 70\% training data, 15\% testing data, and 15\% validation data each time. The thresholds that have the highest average F1 score across all five runs are chosen. When comparing predictions obtained via these thresholds to our human annotations, we obtain an average $R^2$ value of 0.49. The $R^2$ values of each category, as well as the values obtained using alternative hate speech detection approaches, are listed in Table~\ref{tab:annotated_data_performance}.

In addition to using thresholds that maximize F1 scores in the Measuring Hate Speech corpus, we also consider the thresholds that maximize the average $R^2$ value of human annotations. This results in an average $R^2$ value of 0.77 across all categories. However, it is unknown whether the performance of these thresholds will generalize to the subreddits in our sample beyond the random 25 subreddits that we annotated, as the selected thresholds may be overfit to this set of subreddits. Given that it is unclear which set of thresholds will generalize to our entire set of subreddits, we repeat all downstream analyses using the set of thresholds that optimize the average $R^2$ value.

\begin{figure*}[ht]
    \centering
    \includegraphics[width=0.75\textwidth]{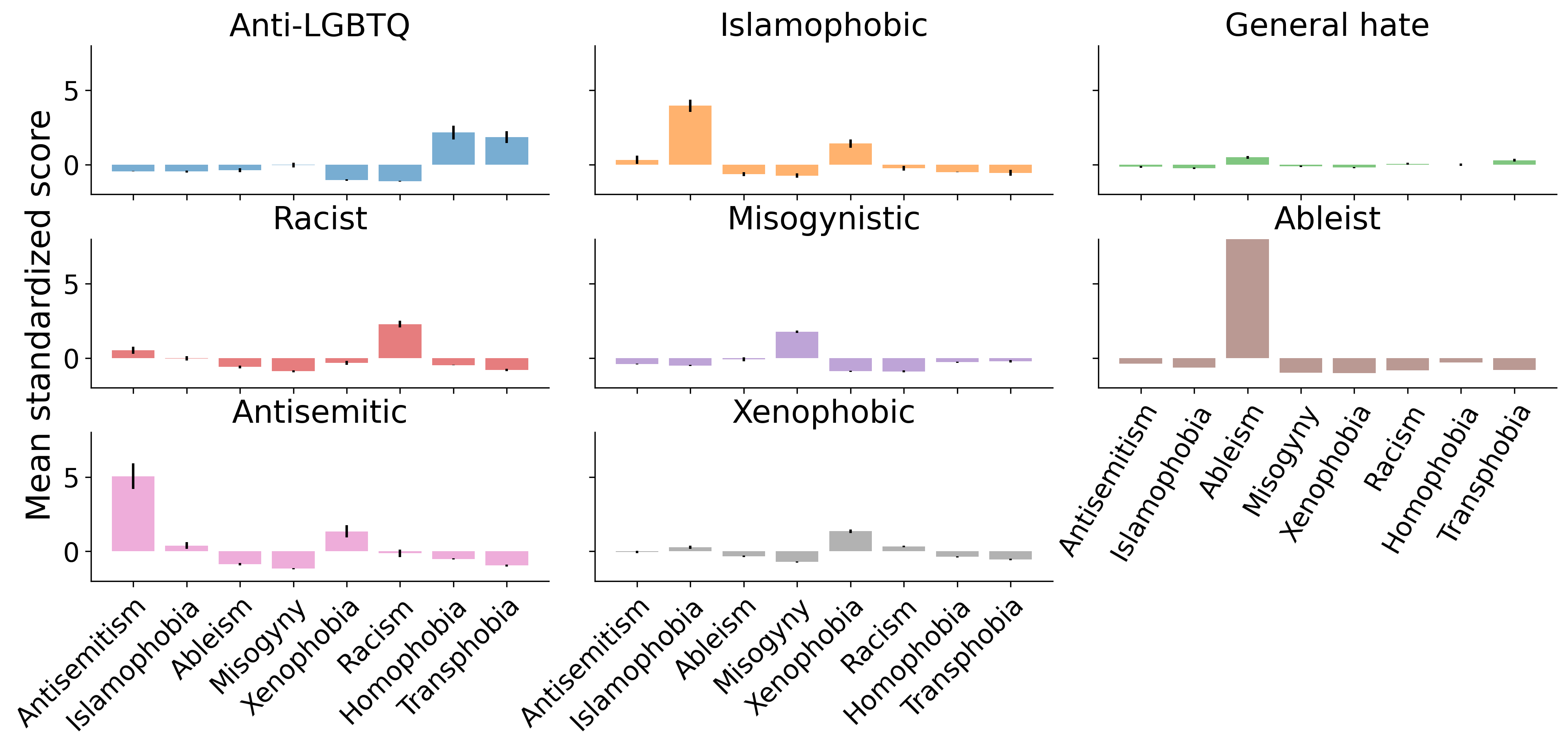}
    \caption{Z-scores of the identities attacked in each cluster. Clusters are titled by the category of hate with the highest z-score (most over-represented).}
    \label{fig:cluster_hate_scores}
\end{figure*}

To separate these subreddits into distinct categories, we employ $k$-means clustering. For each subreddit, we sample and annotate 1K comments and 1K submissions. Each subreddit is assigned a vector of length eight, where each entry represents the proportion of posts targeting at a given identity. Because some categories, misogyny in particular, had a high false positive rate, we transform each estimate into a z-score where z\=0 means the proportion of posts targeting an identity is exactly the mean proportion of posts targeting that identity across all subreddits. We run the $k$-means algorithm for values of $k$ ranging from two to twenty and choose the number of clusters that yield the highest silhouette score (eight clusters). The z-scores for each cluster are shown in Figure~\ref{fig:cluster_hate_scores} and a UMAP plot of the distributions of identities targeted for each subreddit is shown in Figure~\ref{fig:UMAP}. If we vary the model thresholds to maximize the $R^2$ scores and cluster subreddits we find qualitatively similar clusters where the adjusted mutual information score between two sets of clusters is 0.62. We repeat all downstream analyses using this set of clusters and observe qualitatively similar results.


We release the set of subreddits, their clusters, and the average number of posts containing hate speech of each category for each batch of posts, available at the following URL: \url{https://anonymous.4open.science/r/peripatetic-hater-1D26}. Upon publication, we intend to adhere to FAIR principles \cite{fair} by making the data publicly available and easily accessible via GitHub. The data will be in a widely used format (CSV) or accessible via public repositories. 
Caution should be exercised when using the dataset, as several subreddit names contain offensive phrases directed toward marginalized groups. 


We label as \textit{peripatetic users} those individuals who post in an alternative category of hate subreddit at some point after their initial post in a given hate subreddit. Because we seek to explore predictive markers of such peripatetic behavior after users have already participated in an antisocial environment, we restrict our focus to users who post in an alternative category of hate subreddit within six weeks of posting in their first hate subreddit, as we believe early peripatetic behavior will be more closely linked to the text expressed in the initial stages of hate subreddit participation. For robustness, we also show in the Appendix our results when setting the threshold to six months and when not setting any threshold. 

As automated accounts are likely to be ubiquitous across categories of subreddits but are not relevant to the questions at hand, we filter out probable bots from our dataset. Our method follows Schmitz et al. \citeyear{schmitz2023users} and is performed by removing accounts with the following keywords: ``bot,'' ``auto,'' ``transcriber,'' ``gif,'' ``link,'' and ``twitter.'' (Broadly, these qualitatively appear to remove many bot accounts. Organic accounts may include some of these words, such as Twitter, but we find the word ``twitter'' is so rarely used, $\sim100$ accounts, as to not affect results either way). In total, our dataset contains approximately 4.2M users, 226K of whom post in an alternate category of hate subreddit within six weeks of their first post in a hate subreddit, and 3.3M of whom only post in one category of hate subreddit. We report more detailed statistics in Table~\ref{tab:subreddit_cluster_stats}.

\subsection{User Matching}

To understand whether users who initially post in hate subreddits are at risk of becoming engaged in hate subreddits of other types, we employ a process to match each user from a given hate subreddit (a ``joiner'') with a similar ``counterpart'' user who did not engage in a hate subreddit before the joiner, as described in Schmitz et al. \citeyear{schmitz2023users}. While we cannot answer whether joining one hate subreddit will lead a user to join a different type of hate subreddit, matching members of hate subreddits to users who are similar provides a strong baseline for assessing whether there is substantial overlap among the user bases of hate subreddits. Our matching process is summarized as follows: For each hate subreddit, we find non-hate subreddits in which users from the given hate subreddit are most active relative to the total size of the subreddit and then sample potential counterparts from those subreddits. Each user from the hate subreddit is matched to a counterpart using Mahalanobis distance matching. For a given hate subreddit, we only consider users who have posted in the subreddit as their first hate subreddit. The features used to measure the Mahalanobis distance between users are each user's total Reddit ``karma'' (the number of total upvotes minus the number of total downvotes they receive on their posts), their total number of submissions, their total number of comments, their account creation date, and the total number of posts they made in the top 50 non-hate subreddits in which the members of the given hate subreddit were most active. For each counterpart user, all features were measured before the month the joiner posted in the hate subreddit. From these sets of users, we calculate the proportion who post in subreddits of a different type than the given hate subreddit that are joined within our chosen time threshold of six weeks following the users' first post in the given subreddit. We report the proportions for each category of subreddit as a whole. As counterpart users by definition do not participate in the hate subreddit of their treatment counterpart, we record whether or not they become active in a different type of hate subreddit within the same time period as the hate subreddit user to whom they are matched.

In addition to quantifying the level of movement across subreddits through sets of matched users, we also calculate how likely users are to move to a certain category of subreddit given their original category. For each category, we calculate the proportion of peripatetic users who eventually move to each alternate category divided by a preferential attachment null model \cite{barabasi1999} in which peripatetic users join a category proportional to its number of users. This process provides us with a ratio for each pair of origin and destination subreddits, where a value greater than one indicates an unexpectedly greater influx of users from the origin subreddit category, and a value less than one indicates an unexpectedly smaller influx.

\subsection{Peripatetic users and ingroup language}

To explore how ingroup language spreads among subreddits, we use the Sparse Additive GEnerative (SAGE) model of text to extract keywords characteristic of each community \cite{eisenstein2011sparse}. To find the words most characteristic of a set of subreddits, SAGE compares the text from the subreddit to a baseline corpus. For each category of hate subreddit, we obtain the top 300 most characteristic words from SAGE, using the text from the other categories combined as a baseline corpus.

Using the lexicon of ingroup language associated with each category of hate subreddit, we seek to understand how users who post in multiple categories of hate subreddits introduce such language to different types of communities. We identify the hate subreddit category each user first posts and the category they will eventually post in within six weeks.
We then compare the language employed by peripatetic users in the first three days and \textit{prior to} posting in other hate subreddits to the language used by those who exclusively posted within a single category in the first three days. 
This threshold captures users' initial post behavior after adopting the initial subreddit but long before moving to a new hate subreddit category. To check robustness, we also collected all posts from users after joining the first hate subreddit but before posting in any other hate subreddit category. All results from these variations are qualitatively similar, as shown in the Appendix. 

For each combination of origin and destination subreddits, we count the number of terms used from the lexicon of the destination category relative to the number of terms used from all other categories. These are counted separately for peripatetic users and non-peripatetic users. By counting the number of terms associated with each category, we are able to calculate the odds ratio that a peripatetic user employs terms from their destination category within the original category before they posted in the destination category, relative to non-peripatetic users. We assess the statistical significance of differences in language use between users who will eventually become peripatetic and those who will not via Fisher's exact test. As there are some pairs of subreddits with relatively few peripatetic users, which can result in noisy odds ratios, we only report odds ratios for pairs of subreddits in which there are more than 20 peripatetic users that move from the origin subreddit to the destination subreddit.


We also measure the relative increase in ingroup language use from an alternate category after peripatetic users become active in said alternate category. For each peripatetic user, a symmetric window of time is used to measure behavior before and after becoming peripatetic (e.g., if a user posts in a misogynistic subreddit three weeks after posting in a racist subreddit, their behavior in that three weeks is measured as the ``before'' period and their behavior three weeks after posting in the misogynistic subreddit is measured as the ``after'' period).

\subsection{Extracting Topics of Discussion Used by Peripatetic Users}

In addition to understanding how content associated with hate subreddits in which peripatetic users eventually post appears in their postings in their initial community, we are also interested in the general themes expressed by peripatetic users, unbounded by a specific source. To understand these themes, we employ topic modeling using BERTopic \cite{grootendorst2022bertopic}. BERTopic uses pre-trained transformer models to embed text, then clusters those embeddings to form topics. The topics are then summarized using the highest TF-IDF words for each cluster. For each hate subreddit, we fit a topic model using the first three days of posts for each user and prior to joining another subreddit category. As a preprocessing step, we remove URLs and the strings ``[deleted]'' and ``[removed]'' from all posts. Posts containing only URLs or these strings were removed from the dataset entirely. We use the ``outlier reduction'' feature of BERTopic to assign posts that were not assigned to a topic when initially fitting the model to the closest topic based on the cosine similarity of the text embeddings. We then merge the topic models of all subreddits of the same category. After merging the topic models, we count the number of posts in the top 100 topics that peripatetic users and non-peripatetic users employ. For each topic, the odds of a peripatetic user making a post that fits within the given topic are calculated relative to the odds of a non-peripatetic user making a post on the same topic. When comparing the odds ratios of topics, only topics that appear in at least ten percent of the subreddits within the given category are considered, as some topics may come from individual subreddits that are more or less likely to contain peripatetic users. 

\subsection{Predicting Participation In Hate Subreddit Types} \label{sec:model}

To probe the dynamics wherein users join additional categories of hate subreddits, we initially seek to establish whether certain factors are predictive of a user joining a different category of hate subreddit in the first place. Specifically, as we establish how language spreads among the categories of hate subreddits, we are interested in how the text to which users are exposed, and the text that they themselves employ, is indicative of their participation in each type of hate subreddit. To examine the text to which users are exposed, we utilize the posts to which the user replies, as we are confident that the user has read these posts. However, it is important to note that this approach a) excludes a great deal of text that the users likely read within the given subreddit, and b) may reflect self-selection, as the text to which users reply is likely to be especially related to their interests. We frame this as a multi-label classification problem, where we build a neural network trained on the embeddings of each type of text, and have the neural network predict whether a user subsequently joins a subreddit in each category. The ground truth labels are thus binary labels indicating whether a given user posted in each subreddit category within six weeks of participating in their first hate subreddit. The text each user writes or replies to in any subreddit within three days of posting in their initial hate subreddit (including their first post in that subreddit) and prior to joining a new hate subreddit is embedded and fed into our model as input features (we explain embedding below). 

As we are primarily interested in users posting in other categories of subreddits, and not continuing to post in their original category, we do not consider a user making multiple posts in their original category as a positive example. Using this strategy, we compile a dataset of approximately 100k randomly sampled users, 70\% of which was used for training, with 15\% being used each for validation and testing. As some categories (ableist and antisemitic) have very few examples of peripatetic users, we exclude them from the prediction model. In total, the model therefore predicts six categories.

For each user in the dataset, all of the text that they post in their initial subreddit category is chronologically concatenated into the same document, with each post separated by the [SEP] token. The text to which the users reply is separately concatenated in the same way. Each document is embedded into a 768-dimensional vector using Longformer \cite{beltagy2020longformer}. The vectors are then concatenated, resulting in a 1,538-dimensional input layer for the neural network. As some data are missing from Pushshift \cite{gaffney2018caveat}, there are some cases in which all of the posts to which a user replied could not be retrieved; in those cases, a single token (`UNK') was used as a placeholder for the parent text. The category of subreddit in which each user initially posted, as well as the categories in which the users to whom they replied posted prior to their interaction, are input as a one-hot encoding vector into the neural network at the second layer. The final neural network has three layers, with 780 dimensions in the second layer (where 12 of the input features derive from the one-hot encoding describing the initial subreddit and the subreddits of the parent users) and 384 dimensions in the third layer (i.e., each intermediate layer divides the number of features from the text vectors by a factor of two, with additional features in the second layer due to the one-hot encoding). At each hidden layer, a dropout of 0.6 is used, as well as the leaky ReLU activation function. The sigmoid activation function is used at the final output layer, and the model is fit with the binary cross-entropy loss function. We perform the training on an internal computing cluster using an NVIDIA RTX A6000 GPU.

In addition to training a model utilizing both the context and author vectors, we train a model utilizing each vector independently. The same activation functions and dropout values are used, and the number of dimensions is divided by the same factor within each intermediate layer.

Finally, as there is variation in the performance of the model among individual runs, we take random splits of training, validation, and testing. This entails using a different seed for the Python library, \texttt{sklearn}'s \texttt{train\_test\_split} function. After each new split,  
we train the model to calculate the ROC-AUC. This is performed 50 times to determine the mean and standard errors of the ROC-AUC. In addition, we only report performance for each category considering users who did not originate in that category, as we do not consider movement within the same category.

\section{Results}

\textbf{Subreddit Clustering.}

Figure~\ref{fig:cluster_hate_scores} shows the average number of posts predicted as hate speech for each cluster identified by the $k$-means algorithm. From the distributions, it is clear that the clusters mostly align with individual bias categories, such as race or gender. However, some categories exhibit a combination of bias categories: whereas the ``general hate'' category is not particularly high in any bias category, the ``anti-LGBTQ'' category is high in both homophobia and transphobia. With the exception of these categories, we refer to each category of subreddits as the highest-scoring bias category in each subreddit. The counts of subreddits and users per cluster are displayed in Table~\ref{tab:subreddit_cluster_stats}. Figure~\ref{fig:UMAP} shows the distributions of average hate scores for each subreddit plotted in two-dimensional space. Observing the plot, the categories form distinct clusters.

Upon qualitative inspection of the subreddits within each cluster, they seem to thematically fit with the types of hate predicted as most prevalent within each cluster. For example, the anti-LGBTQ cluster contains TERF subreddits such as r/GenderCritical or homophobic subreddits such as r/homophobes. The general hate cluster contains edgy humor subreddits such as r/CringeAnarchy and r/ImGoingToHellForThis, and the xenophobic cluster contains subreddits such as r/The\_Donald which express strong anti-immigrant sentiment. The racist cluster contains subreddits such as r/GreatApes and r/CoonTown, in which anti-Black racism is frequent.

\textbf{RQ1: Users who participate in one type of hate subreddit are likely to subsequently participate in hate subreddits of another type.}

\begin{figure}[ht]
    \centering
    \includegraphics[width=0.9\columnwidth]{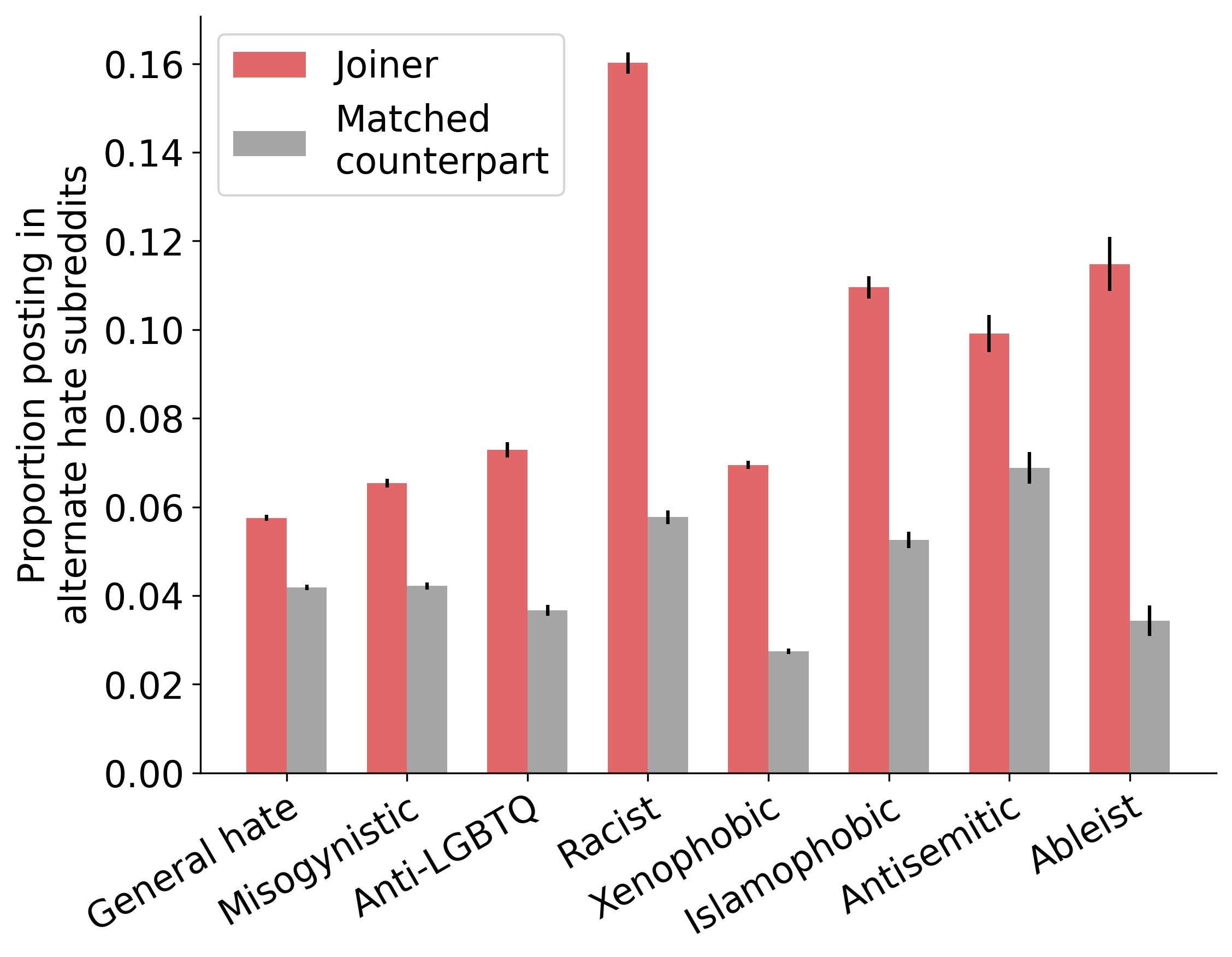}
    \caption{Proportion of users who post in alternative subreddit categories within six weeks after becoming active in their initial hate subreddit category. Error bars represent standard errors.}
    \label{fig:treatment_counterpart}
\end{figure}

Figure~\ref{fig:treatment_counterpart} displays the proportion of newcomers who post in hate subreddits with matched treatment/counterpart pairs. It is apparent that, on average, the users who post in each hate subreddit subsequently participate in many more subreddits of another type than do users who do not post in the given hate subreddit (two times higher depending on the type of subreddit; unweighted mean across all eight categories: 2.1). Using the two-proportion Z test to assess differences between treatment and counterpart groups for each category, we find that the observed results are extremely unlikely to happen by chance, with p-values $< 10^{-5}$ for each result.

\textbf{RQ1: Categories of hate subreddits show preferences for other categories.}

\begin{figure*}[ht]
    \centering
    \includegraphics[width=0.7\textwidth]{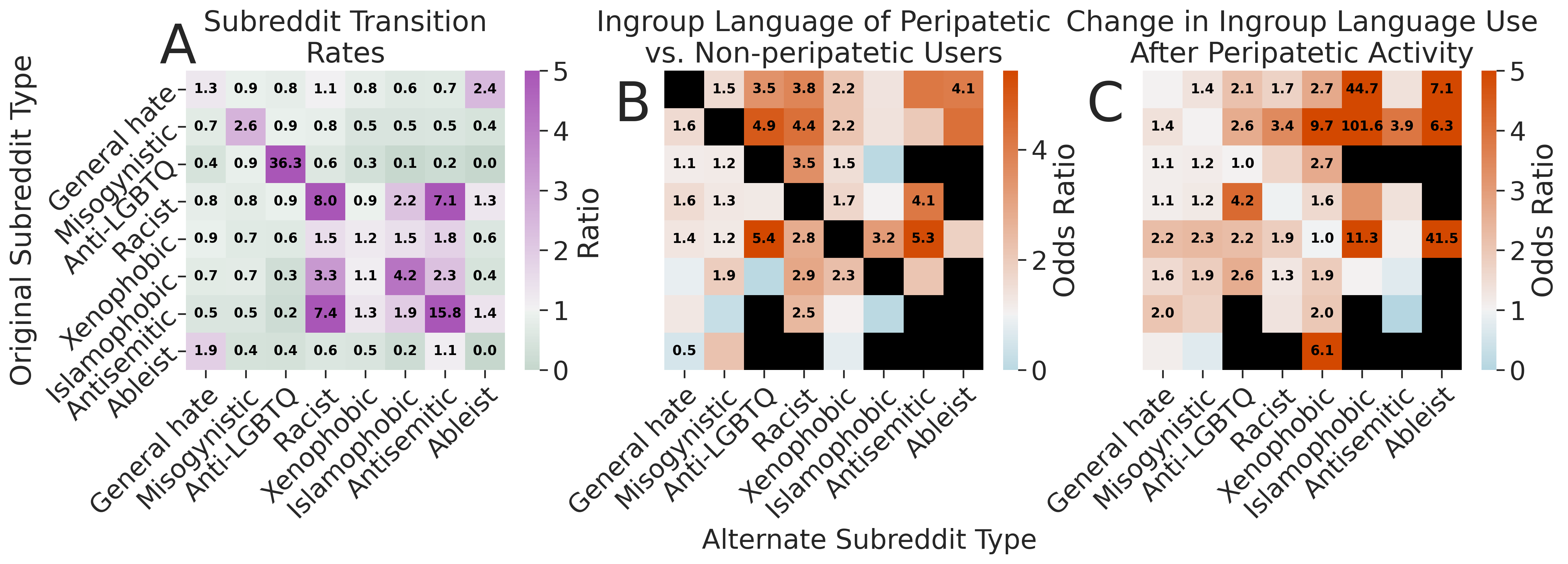}
    \caption{
    Heatmaps representing peripatetic user behavior
    (A) Rate of users from the original subreddit type that subsequently posted in the alternate subreddit. (B) Odds ratios of peripatetic vs. non-peripatetic users using language from the alternate hate subreddit lexicon within the origin subreddit type. (C) Change in use of language from alternate hate subreddit lexicons after posting in alternate hate subreddits. In (B) and (C), cells with numbers in them represent statistical significance (p-values $< 0.05$).}
    \label{fig:transition_matrix}
\end{figure*}

Figure~\ref{fig:transition_matrix}A displays the rate at which members who originate in each subreddit category subsequently post in other categories of subreddits. From the figure, it is apparent that, for most categories, users are more likely than chance to participate again in the category they originated in. Additionally, antisemitic, racist, Islamophobic, and xenophobic subreddits exhibit higher user overlap with each other than expected by chance. Users in the ableist category show a preference for general hate subreddits and vice versa.

\textbf{RQ2: Users who will become peripatetic introduce language associated with alternate hate subreddits to the hate subreddit category in which they initially posted.}

Figure~\ref{fig:transition_matrix}B
displays the odds of peripatetic users posting material, in the category to which they first contributed, that employs language common in the peripatetic user's subsequent subreddit category, relative to the odds that non-peripatetic users employ such language in the first category. The heatmap reveals that, for the most part, the results are statistically significant and are associated with peripatetic users employing more ingroup language associated with the destination subreddit relative to non-peripatetic users. While there are some pairs of categories where we observe the opposite result, these tend not to be statistically significant.

\textbf{RQ2: Peripatetic users discuss specific topics at different rates than non-peripatetic users.}

\begin{figure*}[ht]
    \centering
    \includegraphics[width=\textwidth]{Fig5topics.png}
    \caption{Most frequent topics by category of subreddit. The number of posts within each topic is displayed on the x-axis of each figure. Stopwords were removed from topic representations.}
    \label{fig:topic_frequency}
\end{figure*}

Figure~\ref{fig:topic_frequency} displays the ten most common topics within each category of subreddit. Among these topics, it is apparent that many of the topics are related to the bias categories identified by our hate speech detection model. For example, the top topic in the antisemitic category is represented by the word ``holocaust.'' In the misogynistic category, discussions about feminism, and perceived attractiveness, and topics represented by misogynistic slurs are among the most frequent. In contrast, the general hate category contains discussion about various identities, including topics related to race and feminism. 

\begin{figure*}[ht]
    \centering
    \includegraphics[width=\textwidth]{Fig6odds.png}
    \caption{Topics with the five highest/lowest log-odds ratios of use between peripatetic users and non-peripatetic users. Positive values indicate topics that are over-represented among peripatetic users. Topics with an adjusted p-value greater than 0.05 according to Fisher's exact test are lighter to show statistical insignificance. Stopwords were removed from topic representations.
    }
    \label{fig:topic_ratios}
\end{figure*}

The topics that peripatetic users are most and least likely to address relative to non-peripatetic users are depicted for each category in Figure~\ref{fig:topic_ratios}. In general hate subreddits, topics represented by misogynistic slurs, and those discussing incels or relationships are most highly associated with peripatetic activity, indicating that the users discussing these topics may later move to manosphere subreddits in the misogynistic category. Meanwhile, the topics most positively associated with peripatetic haters in the racist subreddits are comparatively more extreme than those more likely to be employed by non-peripatetic users. In the Islamophobic category, topics represented by misogynistic slurs and the word ``girl'' are more likely to be used by peripatetic users.


\textbf{RQ3: Joining additional hate communities is associated with users developing a broader hate-group lexicon.}
Complementing these results, Figure~\ref{fig:transition_matrix}C shows the odds ratio of ingroup language after users join other hate subreddits (either of the same or a different category). These results show that peripatetic users have a broader hate group lexicon. We caution that this result is not causal, as we do not compare against matched users who do not join additional hate subreddits. These results nonetheless motivate a root concern: not only is joining one hate subreddit harmful \cite{Schmitz2022,schmitz2023users}, but joining additional subreddits presents compounding harm to users and social media. In addition to exploring the change in language use of peripatetic users, we also considered their changes in activity rates. However, we found no significant difference in activity rates before and after users became peripatetic, nor was the change in peripatetic users' activity substantially different from the activity of non-peripatetic users who continued posting in their original category.

\textbf{RQ4: The text that users post, and to which they reply, is predictive of whether they will post in additional categories of hate subreddits.}



\begin{table}[h]
\begin{tabular}{cccc}
\textbf{}         & \multicolumn{3}{c}{\textbf{ROC-AUC}}                             \\ \hline
\textbf{Category} & \textbf{All}          & \textbf{Target}       & \textbf{Context} \\
General hate      & \textbf{.62 +/- .002} & .60 +/- .002          & .58 +/- .002     \\
Misogynistic      & \textbf{.61 +/- .002} & .60 +/- .003          & .56 +/- .003     \\
Anti-LGBTQ        & \textbf{.63 +/- .009} & \textbf{.63 +/- .009} & .58 +/- .009     \\
Racist            & \textbf{.68 +/- .009} & .66 +/- .009          & .60 +/- .010     \\
Xenophobic        & \textbf{.67 +/- .002} & .65 +/- .002          & .58 +/- .003     \\
Islamophobic      & \textbf{.67 +/- .006} & .64 +/- .006          & .62 +/- .006    
\end{tabular}
\caption{Model performance.}
\label{tab:model_predictions}
\end{table}

Table~\ref{tab:model_predictions} lists the performance for our model predicting whether a user will post in a given category of subreddit distinct from their initial category. ``Target'' refers to posts made by the user for whom the prediction is being made, while ``context'' refers to posts to which they reply. 
Uncertainties represent standard errors. For almost all categories of subreddits, the model performs best when supplied with text written by the users and the text to which they replied, compared to either input type in isolation. Notably, the text and metadata of posts users write is more predictive of peripatetic activity than the text and metadata of the users to whom they reply. 

\section{Discussion}

This study provides a comprehensive view of the user dynamics among multiple types of hate communities on Reddit. These results suggest that hate manifested in the current historical moment among English-speaking users of this platform can be categorized into eight dimensions roughly reflecting each major identity from federal definitions of targeted groups. While these results largely corroborate the federal targets of hate, we acknowledge that there could be yet more identities that we did not capture.
How this generalizes to other social media, and how it relates to psychological perceptions of hate, constitute rich avenues for future research. 

Importantly, our results indicate that users who initially participate in one hate community are prone to participate in additional categories of hate communities and that participating in additional categories of hate is associated with a broader hate group lexicon. Using AI tools, we are able to predict which users will join additional hate communities compared to baselines. Part of this predictive power may be because, from the outset, users who will become peripatetic employ more of the language associated with the other hate communities that they will eventually join. Below we discuss the possible implications of these results concerning how hate communities can influence each other online.

\textbf{RQ1: users are likely to participate in multiple communities of hate.} 
Figure~\ref{fig:treatment_counterpart} shows that users who originate in one type of hate community are more likely to participate in an alternate type of hate community than are similar users who did not participate in the original hate community type. This result implies that each type of hate community serves as a ``gateway'' to other hate communities. In other words, \emph{the presence of each type of community could further amplify the growth of the others}, as their user bases appeal to each other. The preferences certain types of hate communities have for each other provides insight into why this may be. For example, racist, xenophobic, antisemitic, and Islamophobic subreddits all have strong user overlap with each other, likely because these types of hate are similar in that they tend to attribute deprecated characteristics to these essences. For example, hate speech directed at Jews and Muslims will often target their perceived race or national origin along with their religious identity.

Future work should explore the reasons why users from each category of hate are more likely to join other hate subreddits -- for example, due to recruitment \cite{russo2024stranger}. Additional research is also needed to determine whether hate group categorizing could help identify previously unknown active hateful communities or clusters, possibly leading to quicker moderation.

\textbf{RQ2: characteristics distinguishing peripatetic and non-peripatetic users.} Given the results displayed in Figure~\ref{fig:transition_matrix}B, it is tempting to conclude that joining additional hate groups leads peripatetic users to import hate terms and ideas when posting in the subreddit in which they first became active. We explored this possibility, but could not find conclusive evidence of the causal direction. Two non-mutually-exclusive explanations therefore seem plausible: 1) users who become peripatetic were interested from the outset in the hate categories of the subreddits that they eventually joined, or 2) users who eventually post in other subreddits may have been consuming content from those subreddits long before they first contributed to them, hence said content may have influenced their posts in the first subreddit that they joined. Regardless of which of these various possibilities applies, given that a) social influence likely occurs, and b) consuming content from, and contributing content to, multiple communities is facilitated by their presence on a single platform, the harm posed by hosting multiple types of hate communities on one platform may be greater than the sum of the harm posed by each community. In this scenario, hate subreddits serve as recruitment venues for each other, both enhancing the growth of their respective user bases and leading to a convergence of hate beliefs over time. Finally, our work on movement complements prior work on understanding the formation of collective identity within these groups \cite{gaudette2021upvoting}. Future work could explore how collective identity affects the likelihood that users become peripatetic.

\textbf{RQ3: Joining additional hate communities may further radicalize users.} Figure~\ref{fig:transition_matrix}C shows an increase in peripatetic users' use of terms associated with the additional categories of subreddits they join. This may imply that users internalize the language of these various dimensions of hate. These features could be a first sign that users may join additional hate groups \cite{schmitz2023users} and is consistent with, although not proof of, theories that propose social networks act as a net to drive individuals to deeper forms of radicalization \cite{borum2011radicalization, maskaliunaite2015exploring, neo2016internet}. Furthermore, other theories explore why people join extreme communities in the first place \cite{mann2023unsorted}. 
Future work is needed to understand whether the motivations for participating in hate communities put forth by these theories play a role in differentiating peripatetic and non-peripatetic hate users. 

\textbf{RQ4: Predicting peripatetic movement.} Table~\ref{tab:model_predictions} illustrates our model's prediction performance for the types of hate subreddits users that will eventually join. The relatively poor performance of the model suggests that stochastic or unobservable factors strongly influence who becomes a peripatetic hater. Nonetheless, for each category, performance is above the baseline, indicating that peripateticism is not a purely stochastic decision. Users are either partly motivated to join based on their past multi-hate category opinions represented in writing, or their writing indicates that they are more vulnerable to being influenced to join other types of hate. This technology is the first step towards understanding users' motivations for joining multiple hate subreddits and sheds light on where interventions designed to reduce the harm of hate subreddits might most productively focus. 

In sum, our exploration of the dynamics of users in Reddit hate communities reveals the existence of both hazards and opportunities. On the one hand, these dynamics suggest that, over time, online hate groups might converge or coalesce, potentially both accelerating their growth and increasing the probability of tangible offline harms. On the other hand, understanding the movement of users among hate communities and how peripatetic users introduce language that cross-fertilizes hate groups may aid in developing avenues for both identification and intervention aimed at inhibiting the growth and coalescence of hate communities, thereby curtailing the spread of hateful beliefs and reducing the risk of offline harms.

\section{Limitations and Future Directions}

While our study effectively categorizes hate subreddits and demonstrates processes of content diffusion across these categories, there are still notable limitations to our work. First, our hate detection methodology could be improved, especially for transphobia. Our results may also vary with alternative models we did not test, and these results may not generalize to non-English hate subreddits. Our model also lacks exhaustive analysis of hyperparameter tuning. While we found that the variations explored so far do not substantially change results, a more exhaustive analysis could improve the model further. Finally, while we filter out likely bots from our dataset, we do not distinguish between users who post in support of a group's ideology and users who employ counterspeech \cite{mathew2019thou}.

Future work can expand on our results by exploring motivations for joining additional hate subreddits in greater detail. For example, a stronger and more explainable predictive model, which explains why the model made the predictions it did, could help platforms develop proactive content moderation strategies by reducing problematic content on social media before it has been disseminated \cite{habib2021proactive}. Additionally, as prior research shows that the proportion of users in a given subreddit who previously posted in banned subreddits is predictive of whether that subreddit will be banned in the future \cite{habib2021proactive}, future work can more closely investigate how peripatetic users may contribute to subreddit bans. 

\section{Acknowledgements}

The authors would like to thank Hillel Cogan, Emily Haddad, Derrick Liu, Daniel Penn, Rafael Arellano, and Dan Faltesek for their valuable work annotating hate speech. Our work is supported by NSF award \#2331722.

\subsection{Paper Checklist}

\begin{enumerate}
    \item  Would answering this research question advance science without violating social contracts, such as violating privacy norms, perpetuating unfair profiling, exacerbating the socio-economic divide, or implying disrespect to societies or cultures?
    Yes, see discussion in Discussion.
  \item Do your main claims in the abstract and introduction accurately reflect the paper's contributions and scope?
Yes.
   \item Do you clarify how the proposed methodological approach is appropriate for the claims made? 
  Yes, see Methods and Discussion.
   \item Do you clarify what are possible artifacts in the data used, given population-specific distributions?
    Yes, see Methods and Discussion. 
  \item Did you describe the limitations of your work?
    Yes, see Discussion and Limitations and Future Directions.
  \item Did you discuss any potential negative societal impacts of your work?
   Yes, see Discussion.
      \item Did you discuss any potential misuse of your work?
     Yes, see Discussion.
    \item Did you describe steps taken to prevent or mitigate potential negative outcomes of the research, such as data and model documentation, data anonymization, responsible release, access control, and the reproducibility of findings?
        Yes, see Methods.
  \item Have you read the ethics review guidelines and ensured that your paper conforms to them?
   Yes.
  \item Did you clearly state the assumptions underlying all theoretical results?
    Yes, see Introduction.
  \item Have you provided justifications for all theoretical results?
   Yes, see Results.
  \item Did you discuss competing hypotheses or theories that might challenge or complement your theoretical results?
    Yes, see Discussion.
  \item Have you considered alternative mechanisms or explanations that might account for the same outcomes observed in your study?
   Yes, see Discussion.
  \item Did you address potential biases or limitations in your theoretical framework?
   Yes, see Limitations and Future Directions.
  \item Have you related your theoretical results to the existing literature in social science?
   Yes, see Related Work.
  \item Did you discuss the implications of your theoretical results for policy, practice, or further research in the social science domain?
   Yes, see Discussion.
  \item Did you state the full set of assumptions of all theoretical results?
   NA
	\item Did you include complete proofs of all theoretical results?
NA
  \item Did you include the code, data, and instructions needed to reproduce the main experimental results (either in the supplemental material or as a URL)?
    Yes, see the Methods section and following URL: \url{https://anonymous.4open.science/r/peripatetic-hater-1D26}
  \item Did you specify all the training details (e.g., data splits, hyperparameters, how they were chosen)?
    Yes, see Methods.
     \item Did you report error bars (e.g., with respect to the random seed after running experiments multiple times)?
    Yes, see Methods.
	\item Did you include the total amount of compute and the type of resources used (e.g., type of GPUs, internal cluster, or cloud provider)?
        Yes, see Methods.
     \item Do you justify how the proposed evaluation is sufficient and appropriate to the claims made? 
    Yes, see Discussion.
     \item Do you discuss what is ``the cost`` of misclassification and fault (in)tolerance?
      Yes, see Discussion.
  \item If your work uses existing assets, did you cite the creators?
    Yes, we use existing data; see Methods.
  \item Did you mention the license of the assets?
   NA
  \item Did you include any new assets in the supplemental material or as a URL?
    Yes, see the following URL: \url{https://anonymous.4open.science/r/peripatetic-hater-1D26}
  \item Did you discuss whether and how consent was obtained from people whose data you're using/curating?
    NA, data are publicly available (not human subject research).
  \item Did you discuss whether the data you are using/curating contains personally identifiable information or offensive content?
    Yes, see Methods.
\item If you are curating or releasing new datasets, did you discuss how you intend to make your datasets FAIR (see \citet{fair})?
    Yes, see Methods.
\item If you are curating or releasing new datasets, did you create a Datasheet for the Dataset (see \citet{gebru2021datasheets})? 
    Yes, see linked repository.
  \item Did you include the full text of instructions given to participants and screenshots?
    NA
  \item Did you describe any potential participant risks, with mentions of Institutional Review Board (IRB) approvals?
    NA
  \item Did you include the estimated hourly wage paid to participants and the total amount spent on participant compensation?
   NA
   \item Did you discuss how data is stored, shared, and deidentified?
   NA
\end{enumerate}

\section*{Appendix}

\begin{table*}[h]
\centering
\begin{tabular}{ccccccc}
\textbf{Cluster} & \textbf{\begin{tabular}[c]{@{}c@{}}Num. \\ subreddits\end{tabular}} & \textbf{Num. users} & \textbf{\begin{tabular}[c]{@{}c@{}}Num.\\ comments\end{tabular}} & \textbf{\begin{tabular}[c]{@{}c@{}}Num.\\ submissions\end{tabular}} & \textbf{\begin{tabular}[c]{@{}c@{}}Avg. comments\\  per user\end{tabular}} & \textbf{\begin{tabular}[c]{@{}c@{}}Avg. submissions\\  per user\end{tabular}} \\ \hline
General Hate     & 58                                                                  & 1,623,808           & 2,716,821                                                        & 434,951                                                             & 1.67                                                                       & 0.27                                                                          \\
Misogynistic     & 32                                                                  & 818,835             & 1,655,450                                                        & 229,105                                                             & 2.02                                                                       & 0.28                                                                          \\
Anti-LGBTQ       & 16                                                                  & 54,341              & 149,444                                                          & 17,290                                                              & 2.75                                                                       & 0.32                                                                          \\
Racist           & 15                                                                  & 55,554              & 140,491                                                          & 15,450                                                              & 2.53                                                                       & 0.28                                                                          \\
Xenophobic       & 35                                                                  & 1,577,444           & 3,445,695                                                        & 477,587                                                             & 2.18                                                                       & 0.30                                                                          \\
Islamophobic     & 7                                                                   & 113,770             & 242,277                                                          & 29,087                                                              & 2.13                                                                       & 0.26                                                                          \\
Antisemitic      & 4                                                                   & 10,174              & 19,433                                                           & 4,957                                                               & 1.91                                                                       & 0.49                                                                          \\
Ableist          & 1                                                                   & 2,770               & 4,183                                                            & 1,519                                                               & 1.51                                                                       & 0.55                                                                         
\end{tabular}
\caption{Subreddit cluster statistics}
\label{tab:subreddit_cluster_stats}
\end{table*}

\begin{table}[h]
\centering
\begin{tabular}{ccc}
\textbf{}         & \multicolumn{2}{c}{\textbf{F1}}          \\ \hline
\textbf{Category} & \textbf{Demux}          & \textbf{BERT}  \\
Racism            & \textbf{0.68 $\pm$ 0.005} & 0.67 $\pm$ 0.001 \\
Misogyny          & \textbf{0.72 $\pm$ 0.001} & 0.71 $\pm$ 0.008 \\
Xenophobia        & \textbf{0.67 $\pm$ 0.017} & 0.65 $\pm$ 0.009 \\
Transphobia       & \textbf{0.31 $\pm$ 0.055} & 0.23 $\pm$ 0.068 \\
Antisemitism      & \textbf{0.62 $\pm$ 0.028} & 0.54 $\pm$ 0.033 \\
Islamophobia      & \textbf{0.63 $\pm$ 0.022} & 0.61 $\pm$ 0.025 \\
Homophobia        & \textbf{0.71 $\pm$ 0.005} & 0.68 $\pm$ 0.006 \\
Ableism           & \textbf{0.69 $\pm$ 0.010} & 0.66 $\pm$ 0.028
\end{tabular}
\caption{Targeted identity prediction performance on the measuring hate speech corpus.}
\label{tab:mhs_performance}
\end{table}

\begin{table*}[h]
\centering
\begin{tabular}{ccccc}
\textbf{}         & \multicolumn{4}{c}{\textbf{$R^{2}$}}                                                                                                                                    \\ \hline
\textbf{Category} & \textbf{Demux (maximize $R^{2}$)} & \textbf{Demux (maximize F1)} & \multicolumn{1}{l}{\textbf{GPT-4 (batches)}} & \multicolumn{1}{l}{\textbf{GPT-4 (individual posts)}} \\
Racism            & \textbf{0.91}                 & 0.83                         & 0.89                                         & 0.60                                                  \\
Misogyny          & \textbf{0.87}                 & 0.50                         & 0.55                                         & 0.29                                                  \\
Xenophobia        & \textbf{0.91}                 & 0.35                         & 0.24                                         & 0.34                                                  \\
Transphobia       & \textbf{0.52}                 & 0.09                         & 0.20                                         & 0.01                                                  \\
Antisemitism      & 0.79                          & 0.79                         & \textbf{0.87}                                & 0.00                                                  \\
Islamophobia      & \textbf{0.65}                 & 0.32              & 0.37                                         & 0.00                                                  \\
Homophobia        & \textbf{0.82}                 & 0.78                         & 0.64                                         & 0.27                                                  \\
Ableism           & \textbf{0.72}                 & 0.29                         & 0.48                                         & 0.24                                                 
\end{tabular}
\caption{Performance of hate speech detection models on annotated subreddit data}
\label{tab:annotated_data_performance}
\end{table*}

\begin{table*}[h]
\centering
\begin{tabular}{cccccc}
\textbf{}     & \textbf{Negative} & \textbf{Disrespectful} & \textbf{Insult} & \textbf{Attack} & \textbf{Hate speech} \\
Negative      & \textbf{-}        & 0.94                   & 0.95            & 0.97            & 0.83                 \\
Disrespectful & \textbf{}         & -                      & 0.99            & 0.93            & 0.87                 \\
Insult        & \textbf{}         &                        & -               & 0.95            & 0.88                 \\
Attack        &                   &                        &                 & -               & 0.84                 \\
Hate speech   &                   &                        &                 &                 & -                   
\end{tabular}
\caption{Correlation matrix of different hate speech detection methods}
\label{tab:correlation_matrix}
\end{table*}

\subsection{Additional robustness checks}

\textbf{GPT-4 Prompting}

In addition to using Demux to detect targeted identities in hate subreddits, we explored the possibility of using GPT-4, as it has been shown to perform well on the task of hate speech detection \cite{kumarage2024harnessing}. For each type of hate speech, we prompt GPT-4 to count the number of posts containing that type of hate speech in batches of 10 with the following string: 

\begin{quote}
Below is a collection of 10 Reddit posts that may or may not be attacking someone because of their [BIAS CATEGORY]. Please respond with the number of these posts that are very clearly [TYPE OF HATE SPEECH]. Do not include anything else in your response:

[LIST OF POSTS]
\end{quote}

Additionally, we test the performance of a prompt designed to detect individual posts:

\begin{quote}
Below is a Reddit post that may or may not be attacking someone because of their [BIAS CATEGORY]. Please respond with ``Yes'' or ``No.'' Is this post very clearly [TYPE OF HATE SPEECH]? Do not include anything else in your response:

[POST]
\end{quote}

Table~\ref{tab:annotated_data_performance} shows the performance of both these approaches on the set of hate subreddits we manually annotated for hate speech. Across the board, GPT-4 performs better when predicting the number of posts containing hate speech rather than predicting whether individual posts contain hate speech. While GPT-4 exhibits comparable performance to Demux, we opt to use Demux as it is less expensive.

\textbf{Subreddit clusters using thresholds that optimize $R^2$.}

As mentioned earlier in the text, we do not know which approach for determining which thresholds to use from our hate speech detection model generalizes better to our broader set of hate communities: using thresholds that maximize $R^2$ values on the data we annotated or using the thresholds that maximize the F1 score on the Measuring Hate Speech corpus. For this reason, we repeat our analyses with both methods. When clustering the subreddits using the thresholds that maximize the $R^2$ value, we find largely the same categories as the thresholds that maximize the F1 score on the Measuring Hate Speech corpus. The only difference is that the ``anti-LGBTQ'' cluster is split up into two clusters: a ``transphobic'' cluster and a ``homophobic'' cluster. Additionally, we noticed many dark humor subreddits appearing in the transphobic cluster (along with TERF subreddits). These dark humor subreddits are classified as general hate subreddits when using the thresholds that maximize the F1 score. Therefore, as the categorization obtained by the threshold that maximized the F1 score fit better with our ethnographic understanding of the subreddits in our dataset, we chose to report the results obtained from this method in the main text of our paper. The results of all analyses are qualitatively similar between the two methods.

\textbf{Varying time thresholds for peripatetic activity.}

\begin{figure}
    \centering
    \includegraphics[width=\columnwidth]{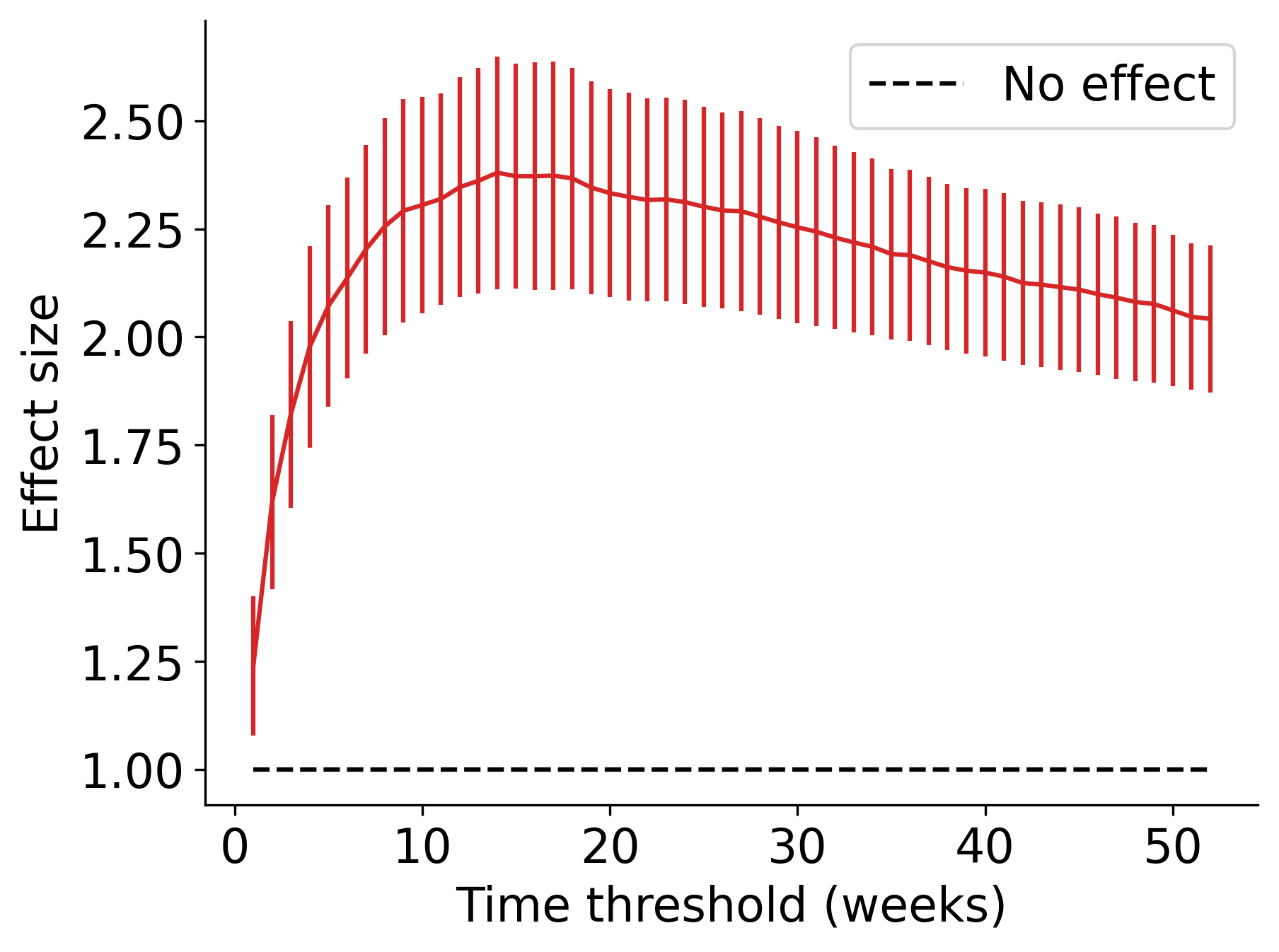}
    \caption{Average effect sizes of Figure~\ref{fig:treatment_counterpart} when considering different time thresholds for peripatetic users. Effect sizes are calculated as the proportion of joiners that post in an alternate category of hate subreddit within the time threshold divided by the proportion of matched counterparts that post in an alternate category of hate subreddit within the time threshold. Error bars represent standard errors.}
    \label{fig:matched_pairs_robustness}
\end{figure}

Figure~\ref{fig:matched_pairs_robustness} shows the average effects of our matched-pairs analysis (Figure~\ref{fig:treatment_counterpart}) when varying the time threshold considered (i.e., what proportion of users post in an alternate hate subreddit within $x$ weeks of posting in their first hate subreddit?). Observing the plot, we can see that the effect size decreases the longer the threshold is. However, when removing the threshold entirely, we find that for all original categories of subreddits, the proportion of joiners who post in an alternate category of hate subreddit remains higher than the proportion of matched counterparts who post in an alternate category. The average effect size when removing the time threshold is 1.4, and all results are statistically significant (p-values $< 10^{-10}$).

\begin{figure*}
    \centering
    \includegraphics[width=0.8\textwidth]{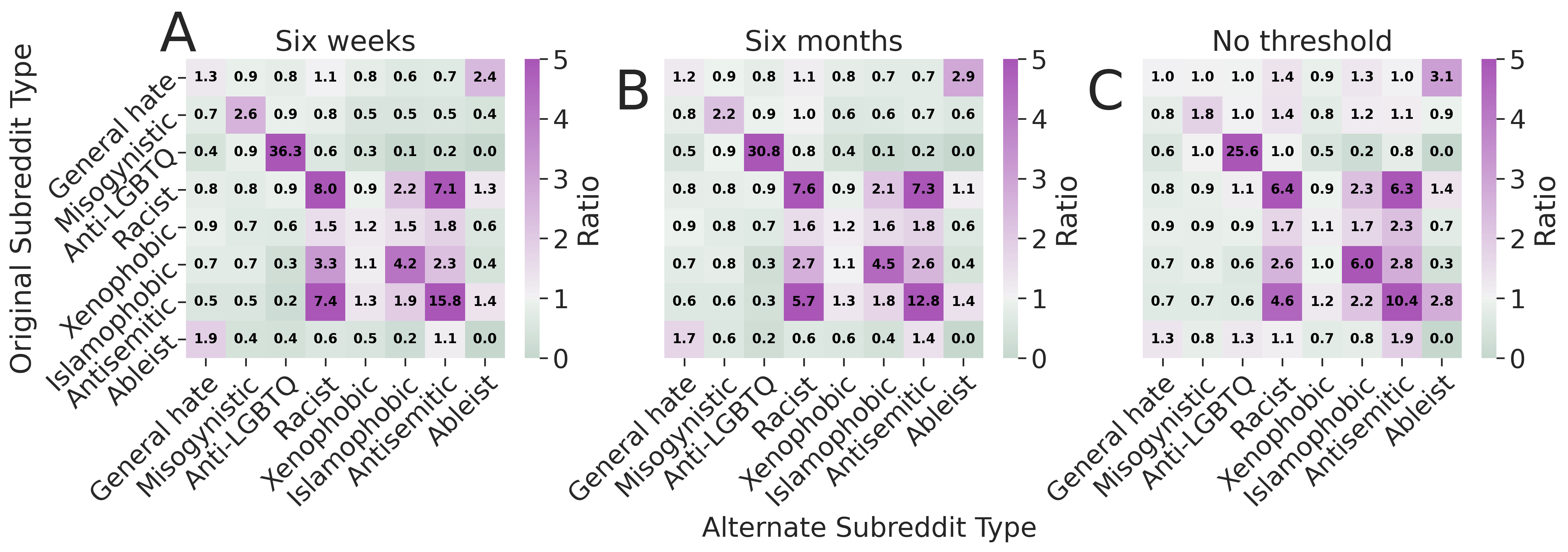}
    \caption{Transition matrices (Figure~\ref{fig:transition_matrix}A) produced when considering different thresholds for considering peripatetic users.}
    \label{fig:robust_transition_matrix}
\end{figure*}

Figure~\ref{fig:robust_transition_matrix} shows the likelihood of a user moving from one category to another within (A) six weeks, (B) six months, and (C) any time at all after initially posting in their first hate subreddit. In all plots, we see similar patterns in which categories of hate subreddits show preferences for each other.

\begin{figure*}
    \centering
    \includegraphics[width=0.8\textwidth]{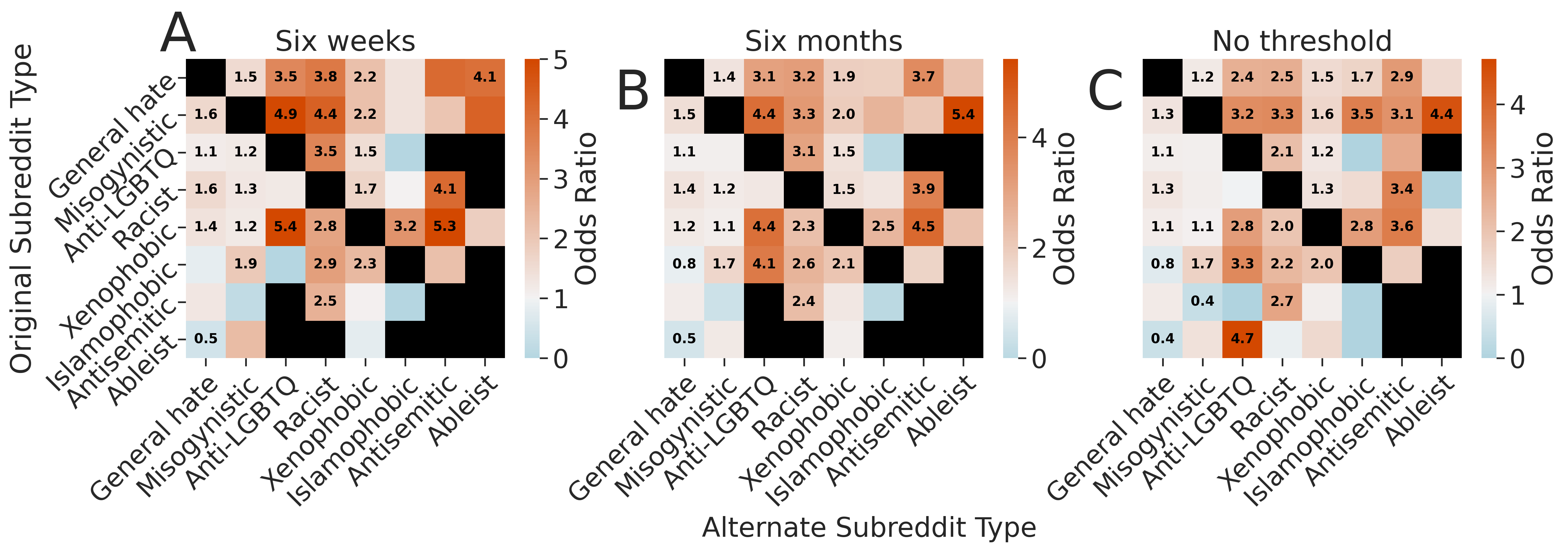}
    \caption{Heatmaps showing relative use of ingroup language by peripatetic vs. non-peripatetic users at different time thresholds. Odds ratios indicate the odds of the use of ingroup language from the alternate category of users who move to the alternate category within the time threshold, relative to the odds of non-peripatetic users using such language.}
    \label{fig:robust_peri_vs_non}
\end{figure*}

\begin{figure*}
    \centering
    \includegraphics[width=0.8\textwidth]{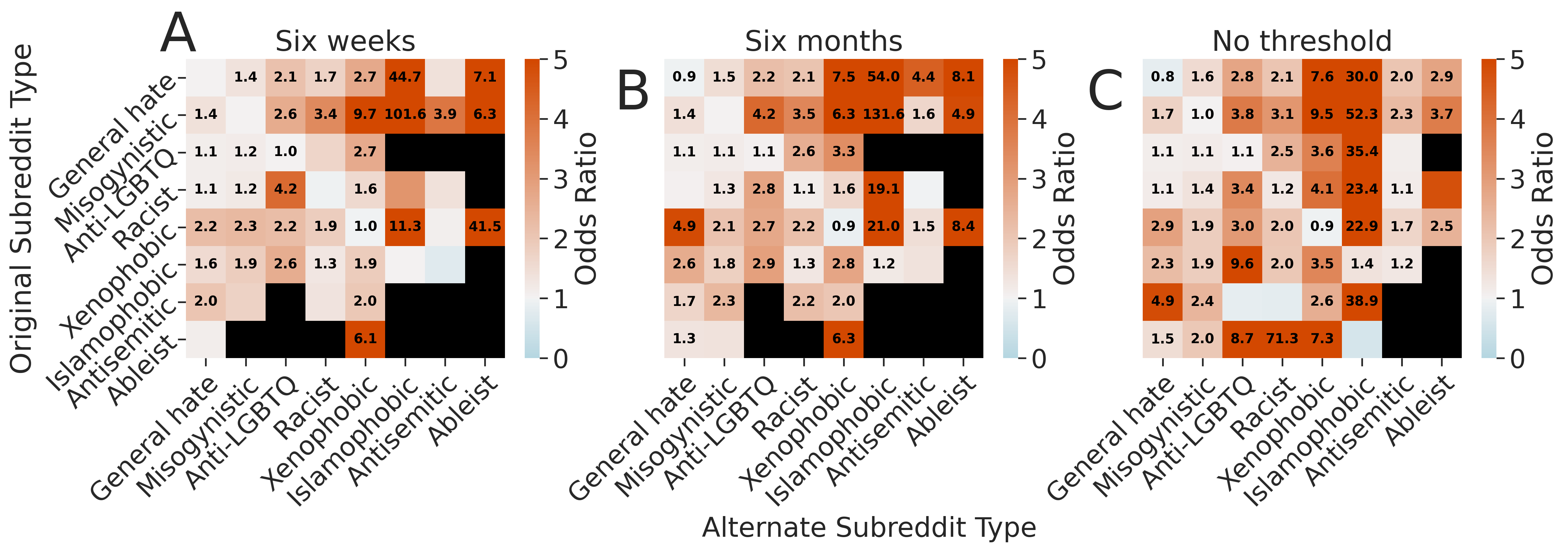}
    \caption{Heatmaps showing the change in use of ingroup language by peripatetic users at different time thresholds. Odds ratios indicate the odds of the use of ingroup language from the alternate category after becoming peripatetic, relative to the odds of the use of ingroup language before becoming peripatetic. Users are considered in the sample if they moved to the alternate category within the time threshold displayed on the subplot.}
    \label{fig:robust_before_after}
\end{figure*}

\begin{figure}
    \centering
    \includegraphics[width=0.8\columnwidth]{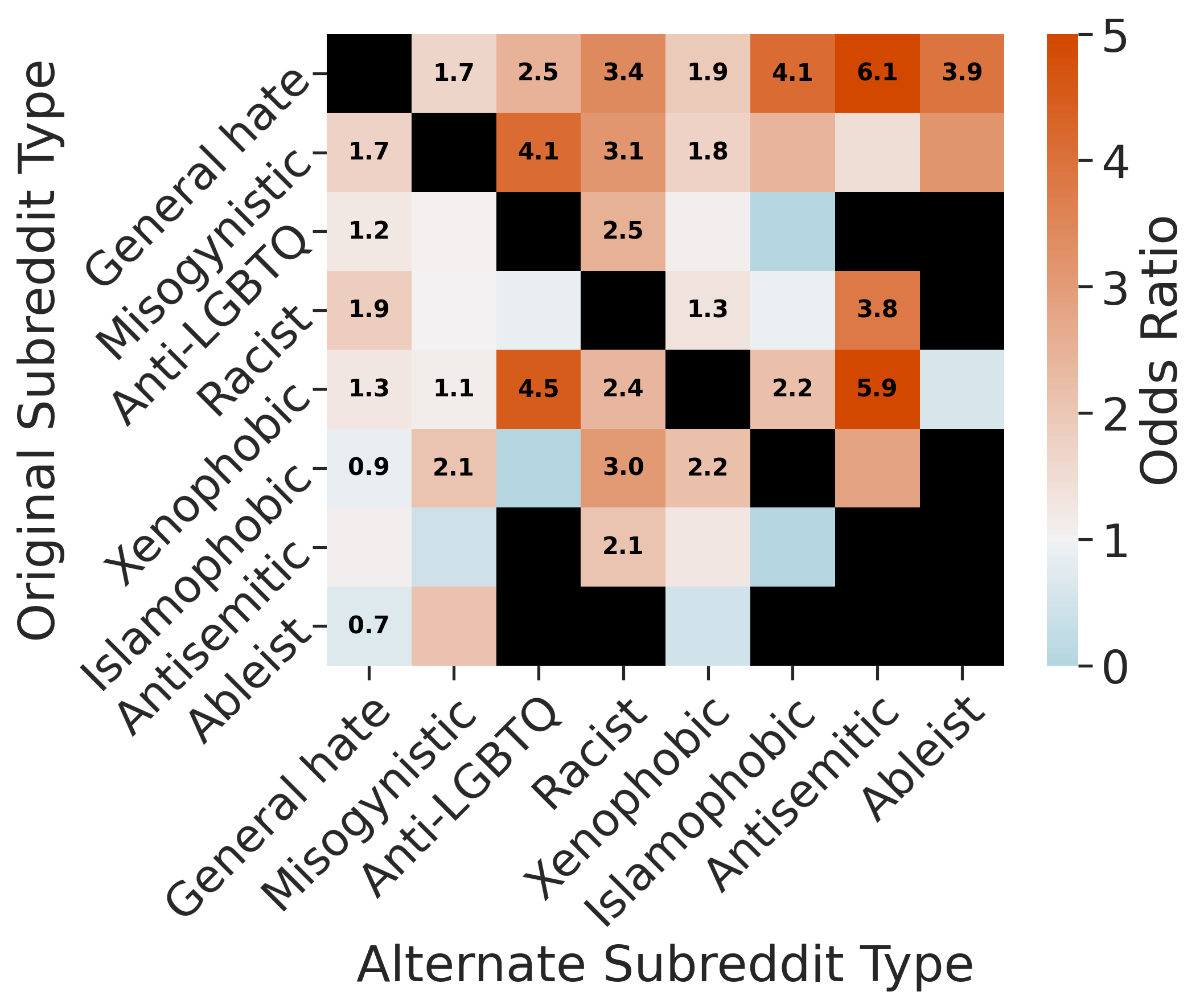}
    \caption{Heatmap showing relative use of ingroup language by peripatetic vs. non-peripatetic users. Unlike Figure~\ref{fig:transition_matrix}B, the odds ratios are calculated from all posts made by non-peripatetic users and all posts made by peripatetic users before moving to an alternate category of hate, instead of only posts made in the first three days.}
    \label{fig:ingroup_language_all_data}
\end{figure}

Figure~\ref{fig:robust_peri_vs_non} shows the results of the analysis displayed in Figure~\ref{fig:transition_matrix}B at varying time thresholds for considering peripatetic users. The effects are mainly positive for all different time thresholds. Figure~\ref{fig:ingroup_language_all_data} shows the results of the same analysis, using posts made at any time before moving to an alternate category of subreddit instead of the first three days (we use the default threshold of six weeks to consider peripatetic users in this case). The results are qualitatively similar to the analysis only considering the first three days of each user's activity. Similarly, Figure~\ref{fig:robust_before_after} shows the results at different time thresholds for the change in ingroup language use after users become peripatetic. We observe similar results regardless of the time threshold.

In addition to showing how our results regarding ingroup language use change when considering all posts instead of only the first three days of each user's activity, we also fit topic models when considering all posts. The top topics are displayed in Figure~\ref{fig:top_topics_all_data}, and the topics with the highest and lowest odds ratios are displayed in Figure~\ref{fig:topic_odds_all_data}. In both plots, many of the topics are represented by similar words to the topic plots in the main text. Though there are also plenty of topics that are different, we nonetheless see similar patterns of topics related to alternate categories of hate being positively associated with peripatetic users (for example, the topic in the racist cluster represented by the words ``gay'' and ``homosexuality'' being used more frequently by peripatetic users).

\begin{figure*}
    \centering
    \includegraphics[width=0.9\textwidth]{Fig12topicsall.png}
    \caption{Most frequent topics per subreddit category when considering all posts made by users prior to posting in an alternate category of hate subreddit. These data include posts made by peripatetic and non-peripatetic users.}
    \label{fig:top_topics_all_data}
\end{figure*}

\begin{figure*}
    \centering
    \includegraphics[width=0.9\textwidth]{Fig13odds.png}
    \caption{Topics most strongly associated with peripatetic activity when considering all posts made by users prior to posting in an alternate category of hate subreddit. These data include posts made by peripatetic and non-peripatetic users. Topics with an adjusted p-value greater than 0.05 are lighter to show statistical insignificance.}
    \label{fig:topic_odds_all_data}
\end{figure*}

\end{document}